\documentclass[prl,showpacs,twocolumn]{revtex4}
\usepackage{mathrsfs}
\usepackage{stmaryrd}
\usepackage{times}
\usepackage{amsmath}
\usepackage{amsfonts}
\usepackage{amssymb}
\usepackage{graphicx}
\usepackage{bm}
\begin{document}

\title{Switching from Crossed Andreev Reflection to Electron Teleportation via Quantum dot }
\author{Jie Liu }
\author{Jian Wang}
\author{Fu-chun Zhang}

\affiliation{ Department of Physics, and Center of Theoretical and Computational Physics, The University of Hong Kong, Hong Kong, China}

\begin{abstract}
We study  electron transport through a normal lead-quantum dot-topological superconductor-quantum dot-normal lead (N-QD-TS-QD-N) junction. Due to the non-local nature of Majorana fermions (MFs) in the topological superconductor, there are two types of single electron transport processes in the junction: crossed Andreev reflection (CAR) and electron teleportation (ET). When the coupling energy of MFs is much larger than the coupling between MFs and QDs, electron can tunnel through topological superconductor either via CAR or ET depending on the energy levels of QDs. For instance, when both energies of QDs (labeled as $\epsilon_1$ and $\epsilon_2$) are
equal to the coupling energy of MFs (denoted as $E_M$), the electron in the left lead can teleport to the right lead via MFs while when $\epsilon_1=-\epsilon_2=E_M$ is satisfied, the electron in the left lead can combine one electron in the right lead to form a cooper pair and tunnel into the topological superconductor via MFs. Since both electron teleportation and crossed Andreev reflection manifest the non-local properties of MFs, they can be used to examine the nature of MFs.

 \end{abstract}

 \maketitle

{\sl Introduction.}   Majorana fermions (MFs) are special type of particles, which are their own antiparticles and obey novel non-Abelian statistics.\cite{ivanov,alicea} Since MFs may non-locally encode a qubit, MFs may form a basic building block in a designed topological quantum computer. Due to its fundamental importance and  potential application in topological quantum computing, it has attracted a great attention to seek the realization of MFs in solid state systems. Indeed, a series of proposals to search for MFs in condensed matter systems have been put forward.\cite{kitaev, read, nayak, sau, fujimoto, sato, alicea2, lut, oreg, potter} One of the promising proposals is that MFs can appear as zero energy end states in 1D superconducting wires by inducing superconductivity
on semiconductor wires with Rashba spin-orbit coupling.\cite{sau, fujimoto, sato} To find the signature of MFs, it was proposed that MFs can induce a local Andreev reflection (AR) which can be used to detect the existence of MFs.\cite{law, wimmer} Soon after this proposal, several groups have fabricated the semiconductor superconducting wire and observed zero bias peaks (ZBP), indicating the existence of MFs.\cite{kou, deng, das1} These observations have  made important first step towards the realization of MFs in solid-state systems. However, the interpretation of the ZBP are often not unique.  An ordinary state which occasionally localizes at the end of the wire could also give rise to a ZBP, and it is difficult to distinguish whether these observed ZBPs are induced by MFs or ordinary localized states. Therefore, it remains highly controversial whether the ZBP has captured the signature of MFs.\cite{jie, altland, piku, kells,tewari,pientka, loss, chill}
\begin{figure}
\centering
\includegraphics[width=3.2in]{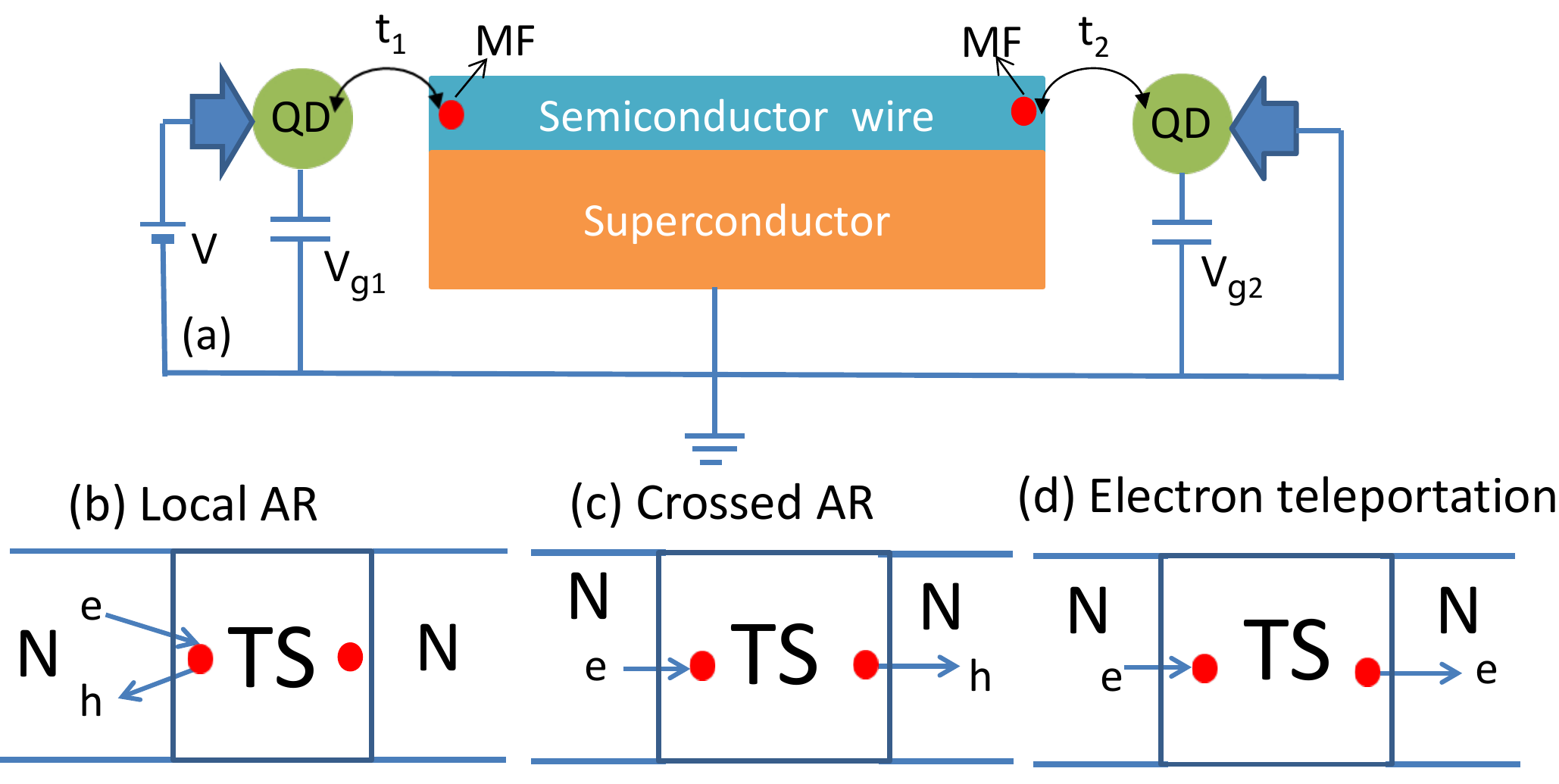}
\caption{ (Color online)  The schematic plot of experimental setup, a well known N-QD-TS-QD-N junction can be used to detect the non-locality of Majorana fermions. }\label{f1}
\end{figure}

To establish the existence of MFs unambiguously, further investigation is clearly needed. The main difference between MFs and ordinary localized states is that MFs are non locally distributed at the two ends of the wire.\cite{kitaev} The origin of non-locality of MFs is due to the fact that a MF can be viewed as half of an ordinary fermion. To define a quantum state with MFs we must consider a pair of non-locally distributed MFs $\gamma_1$ and $\gamma_2$. They combine together to form an ordinary fermionic  state via the relation $f = \gamma_1+i\gamma_2$, while $f$ is an ordinary fermion operator. Since the AR is a local transport process in which an electron tunnels into a superconductor with a hole being reflected back from the same lead as illustrated in Fig. 1(b), the ordinary localized states can lead to the same phenomenon as a MF. The non-local transport of MFs, however, may exclude the possibility contributed from ordinary localized states, hence and should in principle be used to identify MFs. There are many novel non-local transport phenomena. For example, the crossed Andreev reflection (CAR) illustrated in Fig. 1(c) and the electron teleportation (ET) in Fig. 1(d).\cite{nilsson,fu} In the CAR process an electron tunnels into the superconductor from one lead and then tunnels out as a hole at the other lead, while in the ET process an electron tunnels out as an electron instead of a hole at the other lead. If two leads are connected to the ends of a topological superconductor, all three processes may  occur and mix together. In our previous paper, we suggest that the non-local correlation is a good indicator to distinguish these processes.\cite{jie1} However, it will be more interesting to manipulate, control, and measure these non-local processes directly. Given the fact that both non-local processes (CAR and ET) are single electron tunneling events in a lead while the local AR is a two-electron tunneling event, it's very likely that quantum dot (QD) can be used to switch between different transport processes in topological superconductor because a QD can control the number of electrons in each tunneling event by adjusting the coupling strength between the QD and superconductor.\cite{silvano}

In this work, we use two QDs to confine a topological superconductor to form a normal lead-QD-topological superconductor-QD-normal lead (N-QD-TS-QD-N) junction. The experimental setup is depicted as Fig.1(a). Some of the physics in such a junction have been studied in several papers.\cite{tewari1, pei, lu} These papers do reveal many exotic properties of MFs, such as the non-local correlation of MFs, with the aid of QDs. Here, we focus on the transport properties of MFs. Our calculations show that strongly coupled MFs can be confined well within the topological superconductor provided that the QDs are weakly coupled to the MFs. In this situation, two-electron tunneling event (local AR) is largely suppressed and only the single electron tunneling event is dominant. Thus we can neglect the effect of ordinary localized states because they can only enhance the local AR. In addition, an electron can go through the topological superconductor via different single electron tunneling processes depending on the energy level of the QD. When the energy level of both QD (labeled as $\epsilon_1$ and $\epsilon_2$, respectively) are equal to the coupling energy of MFs (denoted as $E_M$), an electron in the left lead can be teleported to the right lead via MFs. When the energy level of one QD $\epsilon_1$ equals to $E_M$ while $\epsilon_2$ equals to $-E_M$, an electron in the left lead may combine with the second electron in the right lead to form a Cooper pair and tunnel into the topological superconductor via MFs. These non-local transport properties certainly can't be induced by ordinary localized states and therefore should be used to identify MFs unambiguously.

{\sl Model and Formalism.}
The Hamiltonian of QD-TS-QD system can be written as follows:
\begin{equation}
H_0  = \sum\limits_{\mathbf{i}} {\varepsilon _{\mathbf{i}} } \mathbf{d}_{\mathbf{i}}^\dag  \mathbf{d}_{\mathbf{i}}  + i\mathbf{E}_{\mathbf{M}} \gamma _1 \gamma _2  +\mathbf{ t}_1 (\mathbf{d}_1^\dag   - \mathbf{d}_1 )\gamma _1  + i\mathbf{t}_2 \gamma _2 (\mathbf{d}_2^\dag   + \mathbf{d}_2 )
\end{equation}
where $\gamma _1 $ and $ \gamma _2$ are the Majorana operators and the parameter $\mathbf{E}_{\mathbf{M}} \propto e
^{-2l/\xi_{0}}\cos(k_{F}l) $ describes the coupling energy between the two MFs,\cite{sarma} where $k_{F}$ is the Fermi momentum and $\xi_{0}$ is the superconducting coherence length. Here $ \mathbf{d}_{1}$ and $ \mathbf{d}_{2}$ are the annihilation operators in the left and right QD respectively and $\mathbf{ t}_1 $ ($\mathbf{ t}_2 $) represents the coupling strength between the left (right) QD and the MF $\gamma_1$ ($\gamma_2$).

\begin{figure}
\centering
\includegraphics[width=3.2in]{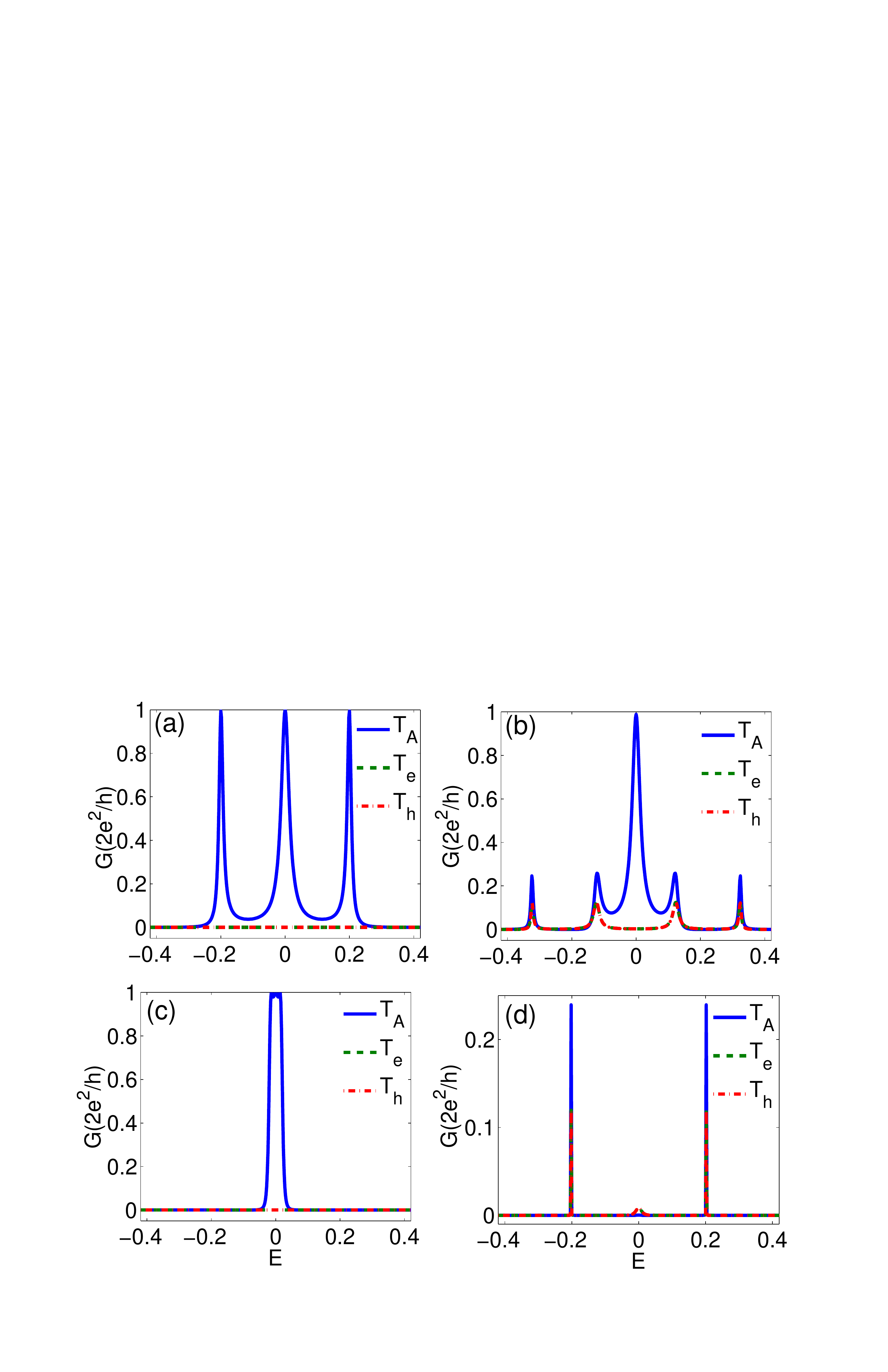}
\caption{ (Color online) (a) and (b) show the conductance as a function of incident energy E in the strong coupling regime with $t_1=t_2=0.1$: (a)  $E_M=0$ means no coupling between the MFs. In this case only the local AR can occur.
Here, $TA$ is the transmission coefficient caused by local AR, $T_e$ is the transmission coefficient of ET, and  $T_h$ denotes the transmission coefficient of CAR. (b)$E_M=0.2$. In this case, the processes of ET and CAR can happen. (c) and (d) show the conductance as a function of incident energy E in the weak coupling regime with $t_1=t_2=0.01$: (c) when $E_M=0$, the local AR still survives. (d) for $E_M=0.2$, all the processes are suppressed due to the weak coupling between QD and MFs except at $E= \pm E_M$. The energies of QD are in line with the fermi level in all the case: $ \varepsilon_1=\varepsilon_2=0$. }\label{f2}
\end{figure}

Since it is convenient to work in the conventional fermion representation, we can transform the Hamiltonian (1) to the following form using these relations $ \gamma _1=f+f^{\dag}$ and $ \gamma _2=-i(f-f^{\dag})$:
\begin{eqnarray}
\tilde{H}_0  &=& H_e+H_{eh},  \nonumber \\
H_e&=&  \sum\limits_{\mathbf{i}} {\varepsilon _{\mathbf{i}} } \mathbf{d}_{\mathbf{i}}^\dag  \mathbf{d}_{\mathbf{i}}  + \mathbf{E}_{\mathbf{M}} f^{\dag}f
+(\mathbf{ t}_1 \mathbf{d}_1^\dag f  +\mathbf{t}_2f \mathbf{d}_2^\dag+H.c.) , \nonumber \\
H_{eh}&=& \mathbf{ t}_1 (\mathbf{d}_1^\dag f^{\dag}  - \mathbf{d}_1 f)+\mathbf{t}_2(f \mathbf{d}_2-f^{\dag}\mathbf{d}_2^\dag), \end{eqnarray}
where the first term $H_e$ is the usual electron Hamiltonian that conserves the number of electron while the second term $H_{eh}$ is the anomalous Hamiltonian which converts two electron into a cooper pair. To study the transport properties of such a system, each QD is connected with a lead. Assuming the tunneling rate of the lead is energy independent, the Green function of the system is easily calculated in the Nambu representation: \cite{lee,sun}
\begin{eqnarray}
G^r  = \left( {E  - \left( {\begin{array}{*{20}c}
   {\tilde{H}_e  + \Sigma _e ^r } & {\tilde{H}_{eh} }  \\
   {\tilde{H}^* _{eh} } & { - \tilde{H}_e ^*  + \Sigma _h ^r }  \\
\end{array}} \right)} \right)^{ - 1},
\end{eqnarray}
where $\tilde{H}_e$ is the matrix representation of $H_e$ with the basis $(d_1^\dag|0\rangle,f^\dag|0\rangle,d_2^\dag|0\rangle)^T$, $\tilde{H}_{eh}$ is the matrix representation of $H_{eh}$, $\Sigma^r_e = \Sigma^r_h = diag(-i\Gamma_L/2,0,-i\Gamma_R/2) $ is the self-energy due to the leads, and $\Gamma_{L(R)}$ is the the linewidth function of the left (Right) leads.  Once the Green function is obtained, we can calculate the current from the left (right) lead directly:\cite{sun}
\begin{subequations}
\renewcommand{\theequation}
{\theparentequation-\arabic{equation}}
\begin{equation}
 I_{L}  = I_{LA}  + I_{Le}  + I_{Lh}, \nonumber
\end{equation}
\begin{equation}
 I_{LA}  = \frac{e}{h}\int {dE } ~ {\rm Tr}[\Gamma _{\mathop{L}} G_{14} ^r  \Gamma _{\mathop{L}}  G_{41} ^a] (f_{\mathop{Le}}  - f_{\mathop{Lh}} ),
\end{equation}
\begin{equation}
 I_{Le}  = \frac{e}{h}\int {dE } ~ {\rm Tr}[\Gamma _{\mathop{L}} G_{13} ^r  \Gamma _{\mathop{R}}  G_{31} ^a] (f_{\mathop{Le}}  - f_{{\mathop{\mathop{ Re}}} } ) ,
\end{equation}
\begin{equation}
 I_{Lh}  = \frac{e}{h}\int {dE } ~ {\rm Tr}[\Gamma _{\mathop{L}} G_{16} ^r \Gamma _{\mathop{R}} G_{61} ^a] (f_{\mathop{Le}}  - f_{{\mathop{Rh}} } ),
\end{equation}
\end{subequations}
where $G_{ij}^r$ is the matrix element of $G^r$ and $G_{ji}^a= (G_{ij}^r)^{\dag}$. The physical meaning of the current is obviously: $I_{LA}$ is the current in the left lead coming from the local AR with the transmission coefficient $T_A={\rm Tr}[\Gamma _L G_{14} ^r \Gamma_L G_{41} ^a]$, while $I_{Le}$ is the current which is contributed by the electron teleportation process with the transmission coefficient $T_e={\rm Tr}[\Gamma _L G_{13} ^r \Gamma _R G_{31} ^a]$, and $I_{Lh}$ is the current due to the contribution of CAR with the transmission coefficient $T_h={\rm Tr}[\Gamma _LG_{16} ^r \Gamma _R G_{61} ^a]$.

{\sl Results and Discussion.}
In the following we will study the transport properties of the QD-TS-QD system. 
As discussed in our previous paper,\cite{jie, jie1} we set the superconducting gap $\Delta$ as our energy unit.
The coupling strength of MFs is then usually on the order of 0.1 and $\Gamma _L/R$ is 
on the order of 0.01. Hence we set $\Gamma _L=\Gamma _R=0.03$ in the following calculation. 
First we consider the case that the QDs are strongly coupled with the MFs with $t_1=t_2=0.1\thicksim E_M$. 
The spectral function of the strong coupling regime has been investigated by Tewari et.al.\cite{tewari1} 
In this situation, the QDs and MFs could form a covalent molecular system. Fig. 2(a) shows the Andreev reflection coefficient $T_A$,  the crossed Andreev reflection coefficient $T_h$ and the electron teleportation coefficient $T_e$ as a function of incident energy E where the energy levels of two QDs $\varepsilon_1=\varepsilon_2=0$ are in line with the fermi level (which is set at superconducting condensate) and the coupling energy of MFs $E_M$ is set to zero (in the case of a long superconducting wire). In this case only the local AR can occur while the nonlocal processes such as the CAR and ET are prohibited which is understandable since zero coupling energy ($E_M$) means no communication between two MFs.
When $E_M$ is nonzero, the non-local process is allowed.
 Fig. 2(b) depicts the results when $E_M = 0.2$ with all other parameters kept the same as in Fig. 2(a). We see that there are four resonant peaks in $T_e$ and $T_h$ as the incident energy E is varied. In addition, $T_A$ also shows four peaks at the same energies. Note that the resonant peak height of $T_A$ is equal to the sum of the corresponding peak height of $T_e$ and $T_h$. This means that for the electron coming from the left QD the probability of reflecting back to the left QD as a hole equals to the probability of tunneling out to the right QD (either as an electron or as a hole). Since all three processes mix together it is hard to distinguish them. Interestingly, $T_A$ versus the incoming
energy also shows a zero bias peak although the energies of MFs are no longer kept at zero due to the coupling between MFs. This is because the energy levels of both QDs are in line with the fermi level. Two electrons can combine together to form a zero energy cooper pair with the help of MFs. If energies of QDs are tuned away from the fermi energy, this process is suppressed. This may explain the reason that the ZBPs can be seen in Das's experiment where the semiconductor wire is so short that the two MFs are strongly coupled to each other.\cite{das1}
    \begin{figure}
\centering
\includegraphics[width=3.2in]{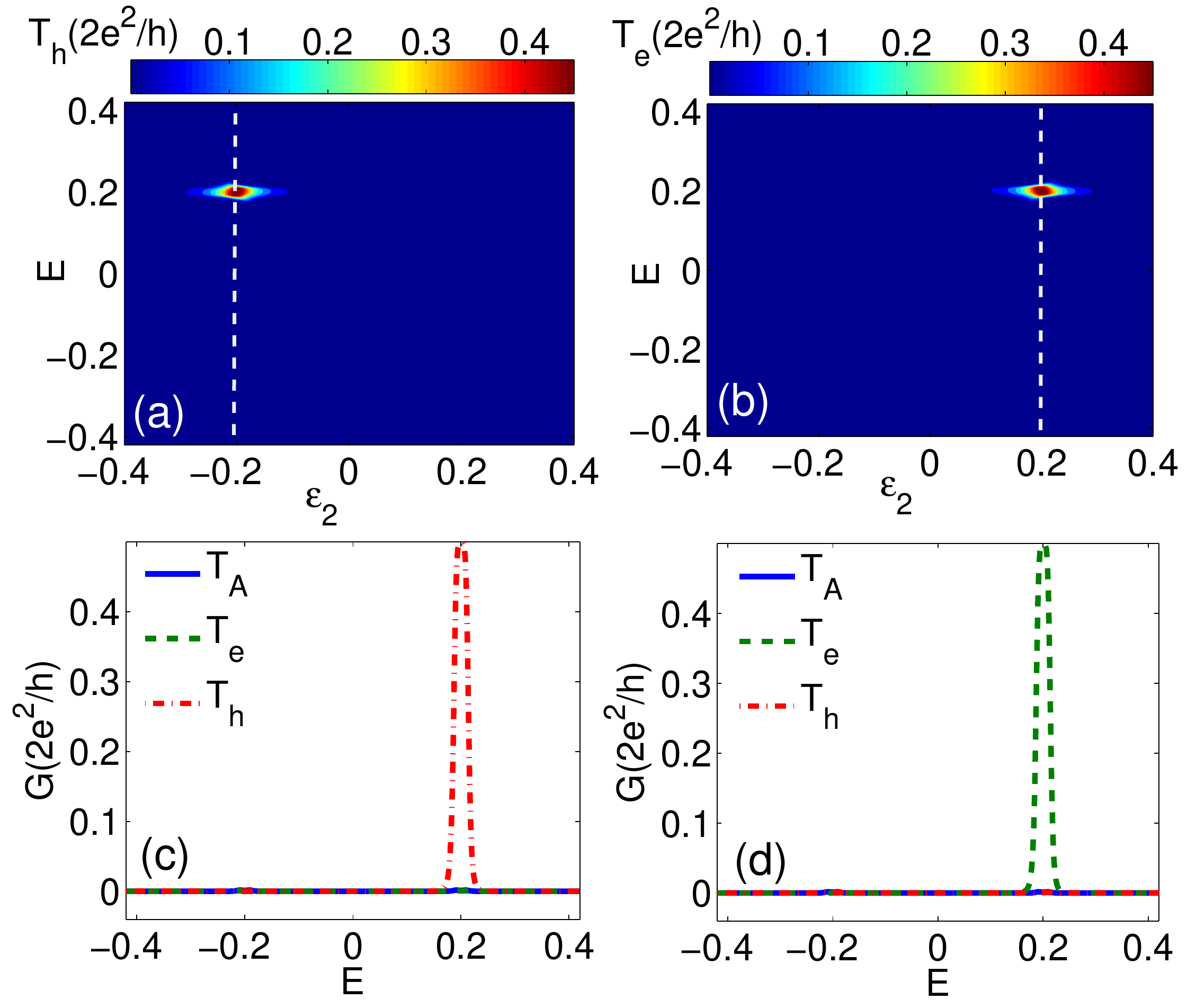}
\caption{ (Color online)   (a) Contour plot of transmission coefficient of CAR $T_h$, as functions of right QD's energy level $\varepsilon_2$ and incident energy E. (b) Contour plot of transmission coefficient of ET  $T_e$. In both cases:
$t_1=t_2=0.01$, and $\varepsilon_1=E_M=0.2$. (c)  $T_A$, $T_e$, $T_h$ as a function of incident energy E with the QD's energies fixed at $\varepsilon_1=-\varepsilon_2=E_M=0.2$ denoted by the dashed lines in (a). In this case, only the CAR survives. (d)  $T_A$, $T_e$, $T_h$ as a function of incident energy E with the QD's energy levels fixed at  $\varepsilon_1=\varepsilon_2=E_M=0.2$ denoted by the dashed lines in (b). The ET is in resonant region while the local AR and CAR are suppressed.}\label{f3}
\end{figure}

From Fig 2. (a) and Fig 2. (b) we can see that the system indeed behaves like a covalent molecule in the strongly coupling regime. Due to the strong coupling the energy levels of the whole system are renormalized and influenced by many parameters such as the energy level of QDs, the coupling strength between MFs and QDs, etc.. In this situation, all three processes, the local AR, CAR and ET, can occur and mix together when $E_M \neq 0$. If the disorder effect is considered in this system, it would further complicate the situation. This is because the ordinary localized states would occur in the presence of disorder and make a contribution to the AR. Thus, one has to suppress the local AR in order to observe the nonlocal processes such as the CAR and ET. A simple way to suppress the local AR is to decrease the coupling strength between MFs and QD. It is known that when the coupling strength between QD and a superconductor is much smaller than the superconducting gap, the two-electron tunneling event is suppressed and only the single electron tunneling event is allowed. Fortunately, both of non-local processes are single electron tunneling events while the local AR are two-electron tunneling event. Thus, with the decreasing of coupling strength between QDs and MFs, the local AR is largely suppressed. Fig. 2 (c) and Fig. 2 (d) show the conductance versus the incident energy E in the weak coupling regime with $t_1=t_2=0.01\ll E_M$. In Fig. 2 (c), $E_M=0$, we can see that the local AR still dominates and the tunneling probabilities of the non-local processes are zero. This is consistent with the previous result that the zero bias peak would remain
at the integer value $2e^2/h$ regardless of the coupling strength between the leads and MFs when $E_M$ is zero. While $E_M$ is not strictly zero, however, the local AR is suppressed. In Fig. 2(d), $E_M=0.2$, we can see that all three processes are almost suppressed except when $E=\pm E_M$ we have resonant peaks with extremely small peak widths.

The most interesting thing is that by tuning the energy level of the QD in the weak coupling regime we can completely suppress the local AR while allowing the processes of CAR and ET. In Fig. 3 (a) and (b) we show the contour plot of $T_h$ and $T_e$ respectively as functions of the right QD's energy $\varepsilon_2$ and the incident energy E. We have set the coupling energy of MFs $E_M = 0.2$, the left QD's energy $\varepsilon_1=E_M$, and the coupling strength between QDs and MFs $t_1=t_2=0.01$. We did not show the contour plot of $T_A$ because it is almost zero. It is very interesting that there is a peak pinned at the location $\varepsilon_1=-\varepsilon_2=E_M=0.2$ in the contour plot of $T_h$ while no such a peak exists at the corresponding location in the contour plot of $T_e$. This means that only the CAR is allowed at this location while other processes are completely suppressed. Similarly, the contour plot of $T_e$ also shows a peak at the location $\varepsilon_1=\varepsilon_2=E_M=0.2$ indicating that the electron teleports from the left QD to the right QD via MFs via resonant tunneling while the other processes are suppressed. To get a clear understanding of these processes, Fig. 3(c) shows the curves of $T_A$, $T_h$, $T_e$ as a function of incident energy E while fixing $\varepsilon_2=-0.2$ (see the white dashed line in Fig. 3(a)). We have also shown in Fig. 3(d) the curves of $T_A$, $T_h$, $T_e$ versus incident energy E with $\varepsilon_2=0.2$ (the white dashed line in Fig. 3(b)). We can see clearly that only two resonant tunnelings occur: resonant CAR at $\varepsilon_1=-\varepsilon_2=E_M=0.2$ and resonant ET at $\varepsilon_1=\varepsilon_2=E_M=0.2$.

In Fig. 3 we demonstrate the switching between two non-local transport processes by tuning the energy levels of QDs. When $\varepsilon_1=\varepsilon_2=E_M$, the current flows from the left lead to the right lead via MFs giving rise to the electron teleportation process while when $\varepsilon_1=-\varepsilon_2=E_M$, the current flow from both two leads to superconductor via MFs due to the cross AR. To understand this unique and interesting feature, we examine the non-locality nature of MFs. It is known that MFs must come in pairs and distribute non-locally at the both ends of the superconducting wire. Since a single MF is just a half fermion, two MFs can combine together to form an ordinary fermonic state with the occupied level and unoccupied level being $E_M$ and $-E_M$, respectively. Hence this state is non-locally distributed at the two ends of the superconducting wire. Thus an electron tunneling into the left end of the wire has probability to appear at the right end with the assistance of MFs. If a lead is attached to the right end of the superconductor, then the electron can tunnel out either as an electron or as a hole. If the MFs are connected by QDs and the coupling strengths between them are small, the states of MFs and QDs are highly localized and in general no electron can tunnel between them except for special energies where resonant tunneling can occur. For instance when  $\varepsilon_1=\varepsilon_2=E_M$ a non-local ET occurs via resonant tunneling
through the occupied state. When $\varepsilon_1=-\varepsilon_2=E_M$, a hole instead of an electron can tunnel out of the right QD with the help of the unoccupied state since the hole energy is in line with the energy level of the right QD showing a non-local CAR. This shows that two non-local processes can be controlled by shifting the energy level of the right QD.

Here we emphasize that the coupling energy of MFs $E_M$ plays an important role in these processes. Due to the coupling between MFs and QDs, energy levels of MFs broaden with a width of order $t_1$ or $t_2$. If $E_M$ is much larger than $t_1$ (the coupling strength between MFs and QDs), then the occupied state and the unoccupied state can be distinguished. In another word, there is no mixing between ET and CAR processes. Fig.4(a) and Fig.4(b) show the contour plots of $T_h$ and $T_e$ respectively as functions of E and $E_M$ where we set $\varepsilon_1=\varepsilon_2=E_M$ and $t_1=t_2=0.01$. While Fig.4(c) and Fig.4(d) show the contour plots of $T_h$ and $T_e$ respectively as functions of E and $E_M$ with $\varepsilon_1=-\varepsilon_2=E_M$. It clearly shows that two processes ET and CAR can occur at the same time when $E_M \sim t_1$ or $ t_2$ while for $E_M \gg t_1, t_2$ only one process can happen. Concerning the coupling energy of MFs, we note that since $\mathbf{E}_{\mathbf{M}} \propto e ^{-l/\xi_{0}}$, $E_M$ can be increased by decreasing the length of superconducting wire. In Kouwenhoven's experiment\cite{kou}, the coherence length $\xi_{0}$ of the superconducting wire is about $250nm$ in a clean system. For a wire whose length is about twice the coherence length of superconducting wire, we have the coupling energy $E_M \sim 0.1\Delta$ which is strong enough to distinguish the two non-local processes. In addition, the thermal broadening width is $k_BT \sim 0.01\Delta$ in Kouwenhoven's experiment. Thus the conductance peak wouldn't be washed out by finite temperature effects.

\begin{figure}
\centering
\includegraphics[width=3.2in]{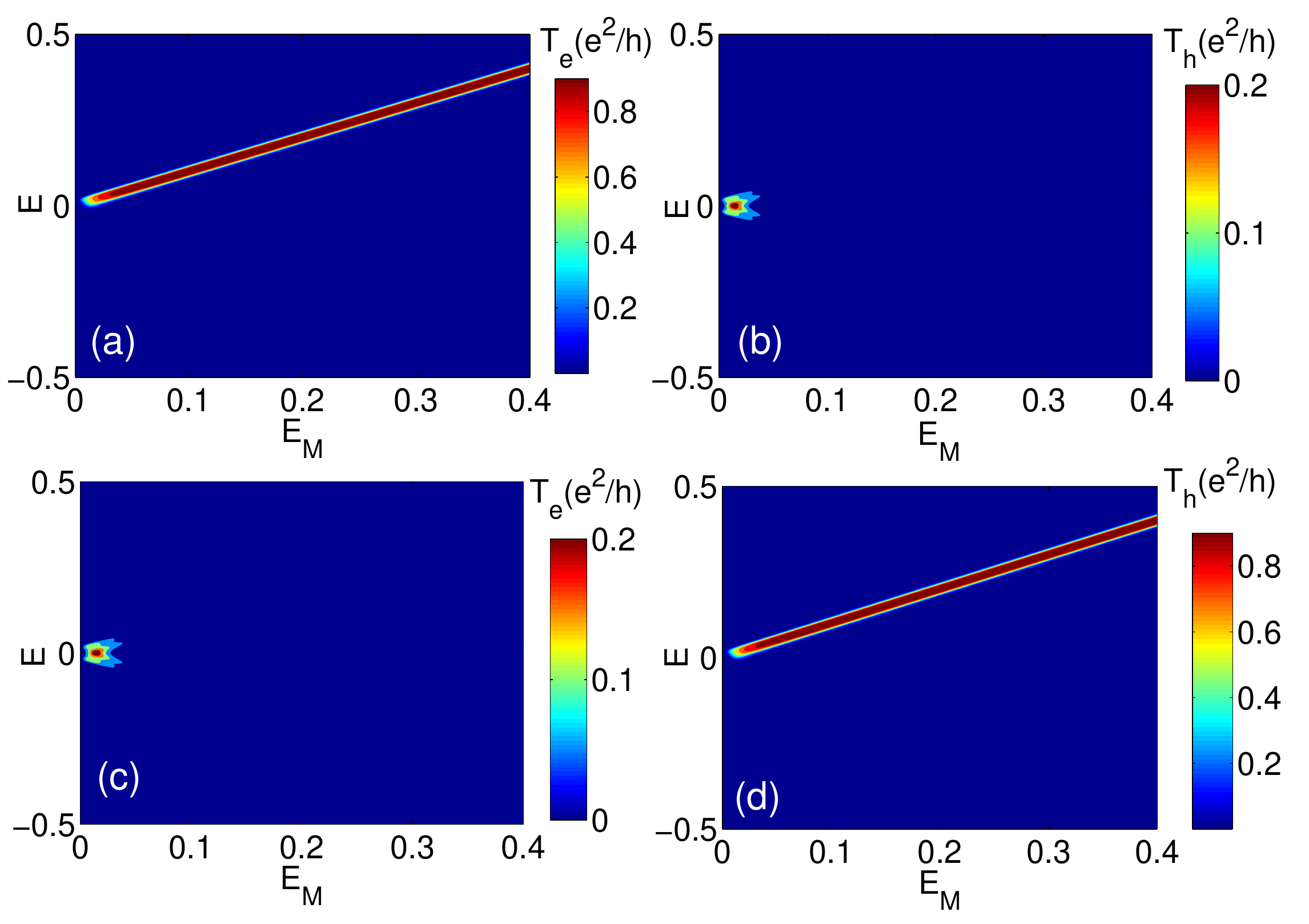}
\caption{ (Color online)   Contour plot of (a)$T_e$ and (b) $T_h$   as a function of MFs' coupling energy $E_M$ and incident energy E.
We fixed the QD's energy level equal to $E_M$: $\varepsilon_1=\varepsilon_2=E_M$. In this case,  only the ET process is allowed.
 Cotour plot of  (c) $T_e$ and (d) $T_h$  as a function of MFs' coupling energy $E_M$ and incident energy E. with the case: $\varepsilon_1=-\varepsilon_2=E_M$. The CAR  is allowed in this case.}\label{f2}
\end{figure}

%

{\sl Conclusion.}  With the aid of QDs, two different types of non-local electron transport processes via MFs in topological superconductor have been investigated. By adjusting energy levels of QDs, switching between these two processes can be achieved, i.e., the resonant ET process will happen when $\varepsilon_1=\varepsilon_2=E_M$ while for $\varepsilon_1=-\varepsilon_2=E_M$, the resonant CAR will occur. Since both ET and CAR manifest the non-local properties of MFs, they can be used to examine the nature of MFs. Importantly, all the conditions for observing resonant ET and CAR can be reached by present technology, we expect that the identification of MFs can be realized in the near future.

{\sl Acknowledgement.}
We thank the insightful discussion with Haiwen Liu, K. T. Law and Hua Jiang.  F. C. Zhang and Jie Liu thank the support of HK RGC GRF grants HKU707211 and AOE/P-04/08. Jian Wang acknowledge the support of HKU 705611P.


\begin{thebibliography} {99}


\bibitem{ivanov} D. A. Ivanov, Phys. Rev. Lett. {\bf 86}, 268 (2001).

\bibitem{alicea} J. Alicea, Y. Oreg, G. Refael, F. von Oppen, M. P. A. Fisher Nature Physics {\bf 7}, 1915 (2011).

\bibitem{kitaev} A. Kitaev, arXiv: cond-mat/0010440 (2000).

\bibitem{read} N. Read and D. Green, Phys. Rev. B {\bf61}, 10267 (2000).

\bibitem{nayak} C. Nayak, S. H. Simon, A. Stern, M. Freedman, S. Das Sarma, Rev. Mod. Phys. {\bf 80}, 1083 (2008).

\bibitem{sau}  J.D. Sau, R.M. Lutchyn, S. Tewari, S. Das Sarma, Phys. Rev. Lett. {\bf 104}, 040502 (2010).

\bibitem{fujimoto} S. Fujimoto, Phys. Rev. B. {\bf 77}, 220501(R) (2008).

 \bibitem{sato}  M. Sato, Y. Takahashi, S. Fujimoto, Phys. Rev. B {\bf 82},134521 (2010).

 \bibitem{alicea2} J. Alicea, Phys. Rev. B {\bf 81}, 125318 (2010).

\bibitem{lut}  R.M. Lutchyn, J.D. Sau, S. Das Sarma, Phys. Rev. Lett. {\bf 105}, 077001 (2010).

 \bibitem{oreg} Y. Oreg, G. Refael, F. von Oppen, Phys. Rev. Lett. {\bf 105}, 177002 (2010).

 \bibitem{potter} A. C. Potter and P. A. Lee, Phys. Rev. B {\bf 83}, 094525 (2011).
 
 \bibitem{law}  K.T. Law, P.A. Lee, T.K. Ng, Phys. Rev. Lett. {\bf 103},237001 (2009).

\bibitem{wimmer} M. Wimmer, A.R. Akhmerov, J.P. Dahlhaus, C.W.J. Beenakker, New J. Phys. {\bf13}, 053016 (2011).

\bibitem{kou} V. Mourik, K. Zuo, S. M. Frolov, S. R. Plissard, E. P. A. M. Bakkers, L.P. Kouwenhoven, Science {\bf 336}, 1003 (2012).

\bibitem{deng} M. T. Deng, C.L. Yu, G.Y. Huang, M. Larsson, P. Caroff, H.Q. Xu, Nano Lett. {\bf12}, 6414-6419 (2012).

\bibitem{das1} A. Das, Y. Ronen, Y. Most, Y. Oreg, M. Heiblum, H.Shtrikman, Nature Physics {\bf 8}, 887 (2012).

\bibitem{jie} J. Liu, A. C. Potter, K.T. Law, P. A. Lee, Phys. Rev. Lett. {\bf109}, 267002 (2012).

\bibitem{altland}  D. Bagrets, A. Altland, Phys. Rev. Lett. {\bf 109}, 227005 (2012)).

\bibitem{piku}  D. I. Pikulin, J. P. Dahlhaus, M. Wimmer, H. Schomerus, C. W. J. Beenakker, New J. Phys. {\bf 14}, 125011 (2012).

\bibitem{kells} G. Kells, D. Meidan, and P. W. Brouwer, Phys. Rev. B {\bf 85},
060507(R) (2012).

\bibitem{tewari} S. Tewari, T. D. Stanescu, J. D. Sau, S. Das Sarma,
Phys. Rev. B {\bf 86}, 024504 (2012).

\bibitem{pientka} F. Pientka, G. Kells, A. Romito, P. W. Brouwer, F. von Oppen, Phys. Rev. Lett. {\bf 109}, 227006 (2012).

\bibitem{loss} D. Rainis, L. Trifunovic, J. Klinovaja, D. Loss  Phys. Rev. B {\bf 87}, 024515 (2013).

\bibitem{chill} H. O. H. Churchill, V. Fatemi, K. Grove-Rasmussen, M. T. Deng, P. Caroff, H. Q. Xu, and C. M. Marcus, Phys. Rev. B {\bf 87}, 241401(R) (2013).

\bibitem{nilsson} J. Nilsson, A. R. Akhmerov, and C. W. J. Beenakker, Phys. Rev. Lett. {\bf 101},120403 (2008).

\bibitem{fu} Liang Fu, Phys. Rev. Lett. {\bf 104}, 056402 (2010).

\bibitem{jie1} Jie Liu, Fu-Chun Zhang, K. T. Law, arXiv: 1212.5879 (2012).

\bibitem{silvano} Silvano De Franceschi, Leo Kouwenhoven, Christian Sch\"{o}nenberger and Wolfgang Wernsdorfer, Nature Nanotechnology {\bf 5}, 703 (2010).

\bibitem{tewari1} S. Tewari, C. Zhang, S. Das Sarma, C. Nayak, and D. H. Lee, Phys. Rev. Lett. {\bf 100}, 027001 (2008).

\bibitem{pei} Peiyue Wang, Yunshan Cao, Ming Gong, Gang Xiong, and Xin-Qi Li, arXiv 1210.5050 (2012).

\bibitem{lu}  H. F. Lu, H. Z. Lu, and S. Q. Shen, Phys. Rev. B {\bf 86}, 075318(2012).

\bibitem{sarma} S. Das Sarma, J. D. Sau, T. D. Stanescu, Phys. Rev. B {\bf 86}, 220506 (2012).

\bibitem{lee}  P. A. Lee and D. S. Fisher, Phys. Rev. Lett. {\bf 47}, 882 (1981); D.S. Fisher and P.A. Lee Phys. Rev. B {\bf 23}, 6851 (1981).

\bibitem{sun}  Qing-feng Sun and X. C. Xie, J. Phys. Condens. Matter. {\bf 21}, 344204 (2009).

%
%


\end{thebibliography}
\end{document}